\documentclass[a4paper,UKenglish,cleveref, autoref, thm-restate]{lipics-v2021}

\title{Inapproximability of Diameter in super-linear time: Beyond the 5/3 ratio}
\titlerunning{Inapproximability of Diameter in super-linear time: Beyond the 5/3 ratio}

\author{\'Edouard Bonnet}{Univ Lyon, CNRS, ENS de Lyon, Université Claude Bernard Lyon 1, LIP UMR5668, France \and \url{http://perso.ens-lyon.fr/edouard.bonnet/}}{edouard.bonnet@ens-lyon.fr}{https://orcid.org/0000-0002-1653-5822}{}

\authorrunning{\'E. Bonnet}

\Copyright{Édouard Bonnet}



\category{}

\supplement{}

\funding{This work was supported by the grant from French National Agency under PRC program (project Digraphs, ANR-19-CE48-0013-01).}

\nolinenumbers 

\EventEditors{Markus Bl\"{a}ser and Benjamin Monmege}
\EventNoEds{2}
\EventLongTitle{38th International Symposium on Theoretical Aspects of Computer Science (STACS 2021)}
\EventShortTitle{STACS 2021}
\EventAcronym{STACS}
\EventYear{2021}
\EventDate{March 16--19, 2021}
\EventLocation{Saarbr\"{u}cken, Germany (Virtual Conference)}
\EventLogo{}
\SeriesVolume{187}
\ArticleNo{47}

\usepackage{xspace}
\usepackage{tikz}
\usepackage{array}
\usepackage{pdfpages}

\usetikzlibrary{fit}
\usetikzlibrary{calc}
\usetikzlibrary{shapes,patterns}
\usepackage{todonotes}

\newcommand{\fov}{\textsc{4-OV}\xspace}
\newcommand{\kov}{\textsc{$k$-OV}\xspace}
\newcommand{\lfov}{\textsc{4-Orthogonal Vectors}\xspace}
\newcommand{\lkov}{\textsc{$k$-Orthogonal Vectors}\xspace}

\newcommand{\dist}{d}
\newcommand{\diam}{\text{diam}}

\newcommand{\ind}{\text{ind}}

\newcommand{\abc}{\text{ABC}\xspace}
\newcommand{\dcb}{\text{DCB}\xspace}

\newcommand{\adx}{\text{AD}_X\xspace}
\newcommand{\ady}{\text{AD}_Y\xspace}

\newcommand{\ab}{\text{AB}\xspace}

\newcommand{\dc}{\text{DC}\xspace}

\definecolor{g1}{rgb}{0,0,1}
\definecolor{g2}{rgb}{0.2,0,0.8}
\definecolor{g3}{rgb}{0.4,0,0.6}
\definecolor{g4}{rgb}{0.6,0,0.4}
\definecolor{g5}{rgb}{0.8,0,0.2}
\definecolor{g6}{rgb}{1,0,0}

\begin{document}

\maketitle

\begin{abstract}
  We show, assuming the Strong Exponential Time Hypothesis, that for every $\varepsilon > 0$, approximating directed \textsc{Diameter} on $m$-arc graphs within ratio $7/4 - \varepsilon$ requires $m^{4/3 - o(1)}$ time.
  Our construction uses non-negative edge weights but even holds for sparse digraphs, i.e., for which the number of vertices $n$ and the number of arcs $m$ satisfy $m = \Tilde{O}(n)$.
  This is the first result that conditionally rules out a near-linear time $5/3$-approximation for a variant of~\textsc{Diameter}. 
\end{abstract}

\section{Introduction}\label{sec:intro}

The diameter of a graph is the largest length of a shortest path between two of its vertices.
We denote by \textsc{Diameter} the algorithmic task of computing the diameter of an input graph.
We will sometimes prefix \textsc{Diameter} by the adjectives \emph{undirected/directed} specifying if edges can be oriented (i.e., if they can be arcs), and \emph{unweighted/weighted} specifying if non-negative edge weights can be used.
By default, we will assume that both edge orientations and edge weights are allowed.
To be clear, the diameter in a digraph (or directed graph) is the maximum taken over all \emph{ordered} pairs of vertices $(u,v)$ of the distance from~$u$ to~$v$.
Note that it is very possible that the pair $(u,v)$ realizes the distance of the diameter, while there is a much shorter path from $v$ to $u$ (perhaps just an arc). 

There is an active line of work aiming to determine the best running time for an algorithm approximating (variants of) \textsc{Diameter} within a given ratio (see for instance the survey of Rubinstein and Vassilevska Williams \cite{Rubinstein19}).
We focus here on sparse graphs, for which the number of edges $m$ and the number of vertices $n$ verify $m = \Tilde{O}(n)$, where $\Tilde{O}$ suppresses the polylogarithmic factors.\footnote{Throughout the paper we adopt the convention that $n$ denotes the number of vertices and $m$, the number of edges of a given graph.}  
There is an exact algorithm running in time $\Tilde{O}(n^2)$ by computing $n$ shortest-path trees from every vertex of the graph.
There is also a folklore $2$-approximation running in time $\Tilde{O}(n)$ by computing a shortest-path tree from an arbitrary vertex and outputting the largest distance found.
There are an $\Tilde{O}(n^{3/2})$ time $3/2$-approximation for weighted directed \textsc{Diameter}~\cite{Aingworth99,Roditty13,Chechik14}, and for every non-negative integer $k$, an $\Tilde{O}(n^{1+\frac{1}{k+1}})$ time $(2-2^{-k})$-approximation for weighted \emph{undirected} \textsc{Diameter}~\cite{Cairo16}.
In dense graphs these four algorithms take time $\Tilde{O}(mn)$, $\Tilde{O}(m)$, $\Tilde{O}(m^{3/2})$ and $\Tilde{O}(mn^{\frac{1}{k+1}})$, respectively.

There are two competing criteria: minimizing the approximation factor, which is in that case between 1 and 2, and minimizing the exponent of the running time, also a real number between 1 and 2.
We now know that the points $(1,2)$, $(\frac{3}{2},\frac{3}{2})$, and $(2,1)$ are feasible for the more general variant of~\textsc{Diameter}.
The question is whether these algorithms can be improved or if conditional lower bounds can be provided instead.
The paper is about the latter, so we will now briefly present the relevant framework of fine-grained complexity, as well as its known consequences for~\textsc{Diameter}.   

\paragraph*{Fine-grained complexity}

The program of fine-grained complexity aims to match fine-grained algorithms (where the precise running time matters more than the membership to some robust complexity class) with tight conditional lower bounds under well-established assumptions.
These assumptions are said \emph{problem-centric}.
They rely on the fact that we have been collectively unable to ``meaningfully'' improve over the brute-force or textbook algorithms for some important problems.
Then perhaps such improvements are impossible, or at least they are currently out of reach.
A fine-grained reduction from one of these central problems to our problem of interest $\Pi$ tells us that improving on $\Pi$ would result in a major breakthrough.

The three main hypotheses are based on \textsc{SAT}, \textsc{3-SUM}, and \textsc{All-Pairs Shortest-Paths}.
One might think that \textsc{All-Pairs Shortest-Paths} is a better starting point for a reduction to~\textsc{Diameter}.
Surprisingly it happens that \textsc{SAT} is.
We will now focus on conditional lower bounds for~\textsc{Diameter}, so we will only define the hypothesis based on \textsc{SAT}.
For more on fine-grained complexity, we refer the interested reader to the survey of Vassilevska Williams~\cite{WilliamsSurvey}.

The Strong Exponential Time Hypothesis (SETH, for short) asserts that for every $\varepsilon > 0$, there is an integer $k$ such that \textsc{$k$-SAT} cannot be solved in time $(2-\varepsilon)^n$ on $n$-variable instances~\cite{Impagliazzo01}.
The first SETH-based lower bound for a polynomial-time solvable graph problem was precisely on unweighted undirected \textsc{Diameter}~\cite{Roditty13}.
The authors show that, unless the SETH fails, any $(3/2 - \delta)$-approximation for sparse unweighted undirected \textsc{Diameter}, with $\delta > 0$, requires time $n^{2 - o(1)}$.
Backurs et al.~\cite{Backurs18} show under the same assumption that, for every $k \geqslant 3$, any $((5k-7)/(3k-4) - \delta)$-approximation for sparse unweighted directed \textsc{Diameter}, with $\delta > 0$, requires time $n^{1+\frac{1}{k-1} - o(1)}$, and that any $(5/3 - \delta)$-approximation for sparse weighted undirected \textsc{Diameter} requires time $n^{3/2 - o(1)}$.
Li~\cite{Li20} improves on these results showing that, unless the SETH fails, any $(5/3 - \delta)$-approximation for unweighted undirected \textsc{Diameter} requires time $n^{3/2 - o(1)}$.

Since a $5/3$-approximation of \textsc{Diameter} running in near-linear time was consistent with the current knowledge, even in weighted directed graphs, Rubinstein and Vassilevska Williams \cite{Rubinstein19} and Li~\cite{Li20} ask for such an algorithm or some lower bounds with a ratio closer to~2.
We give an evidence that, at least for weighted directed graphs, no such algorithm is possible.
More precisely, our contribution is the following.

\begin{theorem}\label{thm:main}
  Unless the SETH fails, for any $\varepsilon > 0$, $(7/4 - \varepsilon)$-approximating \textsc{Diameter} on directed $n$-vertex $\Tilde{O}(n)$-edge graphs where all the edge weights are non-negative integers requires $n^{4/3 - o(1)}$ time. 
\end{theorem}

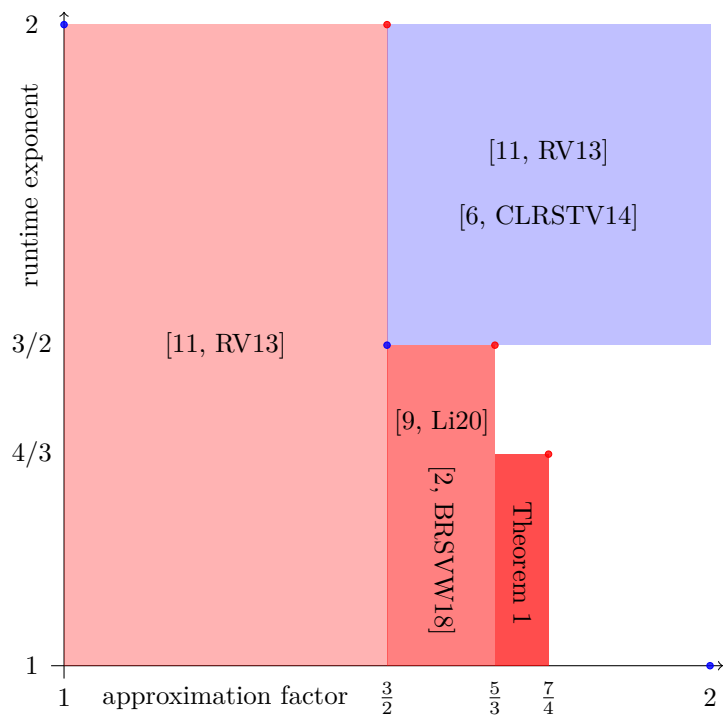
\begin{figure}[h!]
  \centering
  \begin{tikzpicture}
    \def\s{8.5}
    \def\xb{1}
    \def\xe{2}
    \def\yb{1}
    \def\ye{2}
    \def\h{0.02}
    \draw[->] (\xb * \s, \yb * \s - \h * \s) --  (\xb * \s, \ye * \s + \h * \s) ;
    \draw[->] (\xb * \s - \h * \s, \yb * \s) --  (\xe * \s + \h * \s, \yb * \s) ;
    \foreach \i/\j in {1/1,1.33/{4/3},1.5/{3/2},2/2}{
      \node at (\xb * \s - 0.05 * \s, \i * \s) {\j} ;
    }
    \foreach \i/\j in {1/1,1.5/{$\frac{3}{2}$},1.667/{$\frac{5}{3}$},1.75/{$\frac{7}{4}$},2/2}{
      \node at (\i * \s, \yb * \s - 0.05 * \s) {\j} ;
    }
 
    \node at (1.25 * \s, \yb * \s - 0.05 * \s) {approximation factor} ;
    \node at (\xb * \s - 0.05 * \s, 1.75 * \s) {\rotatebox{90}{runtime exponent}} ;
    \fill[red,opacity=0.3] (\xb * \s, \yb * \s) -- (1.5 * \s, \yb * \s) -- (1.5 * \s, \ye * \s) -- (\xb * \s, \ye * \s) -- cycle ;
    \node at (1.25 * \s, 1.5 * \s) {\cite[RV13]{Roditty13}} ;

    \fill[red,opacity=0.5] (1.5 * \s, \yb * \s) -- (1.667 * \s, \yb * \s) -- (1.667 * \s, 1.5 * \s) -- (1.5 * \s, 1.5 * \s) -- cycle ;
    \node at (1.585 * \s, 1.38 * \s) {\cite[Li20]{Li20}} ;

    \node at (1.585 * \s, 1.18 * \s) {\rotatebox{-90}{\cite[BRSVW18]{Backurs18}}} ;

    \fill[red,opacity=0.7] (1.667 * \s, \yb * \s) -- (1.75 * \s, \yb * \s) -- (1.75 * \s, 1.33 * \s) -- (1.667 * \s, 1.33 * \s) -- cycle ;
    \node at (1.71 * \s, 1.16 * \s) {\rotatebox{-90}{\cref{thm:main}}} ;

    \fill[blue,opacity=0.25] (1.5 * \s, 1.5 * \s) -- (1.5 * \s, \ye * \s) -- (\xe * \s, \ye * \s) -- (\xe * \s, 1.5 * \s) -- cycle ;
    \node at (1.75 * \s, 1.8 * \s) {\cite[RV13]{Roditty13}} ;
    \node at (1.75 * \s, 1.7 * \s) {\cite[CLRSTV14]{Chechik14}} ;

    \foreach \i/\j/\c in {1/2/blue,2/1/blue,1.5/2/red,1.5/1.5/blue,1.667/1.5/red,1.75/1.33/red}{
      \draw[\c] (\i * \s, \j * \s) circle [radius=1.2pt] ;
      \fill[\c,opacity=0.8] (\i * \s, \j * \s) circle [radius=1.2pt] ;
    }
  \end{tikzpicture}
  \caption{Approximability of sparse directed~\textsc{Diameter}.
    The blue region is feasible, as witnessed by algorithms at the bottom-left corners (blue dots).
    The three algorithms support non-negative edge weights.
  The red regions would refute the SETH, as witnessed by reductions at top-right corners (red dots).
  The lower bounds in~\cite{Roditty13,Li20} even hold for the sparse unweighted undirected~\textsc{Diameter}, and the one in~\cite{Backurs18}, for sparse weighted undirected~\textsc{Diameter}, while~\cref{thm:main} uses edge weights and orientations.
  }
  \label{fig:results}
\end{figure}

\cref{fig:results} summarizes what was known for directed~\textsc{Diameter} and where~\cref{thm:main} fits.

\paragraph*{Techniques}

Like all the \textsc{Diameter} lower bounds (see also~\cite{WilliamsSurvey,Rubinstein19}), we reduce from \lkov.
In this problem, given a set of $N$ $0,1$-vectors of dimension $\ell$, one looks for $k$ vectors such that at every index, at least one of these $k$ vectors has a 0 entry. 
Unless the SETH fails, \lkov requires time $N^{k-o(1)}$~\cite{Williams05}, even when $\ell$ is polylogarithmic in $N$.

Here we will more precisely reduce from \lfov (\fov, for short).
We want to build a digraph on $\Tilde{O}(N^3)$ vertices and arcs, with diameter 7 if there is an orthogonal quadruple (that is, a solution to the \fov instance), and diameter 4 otherwise.
Following a reduction to $ST$-\textsc{Diameter}\footnote{where one seeks the length of a longest shortest path from a vertex of $S$ to a vertex of $T$} by Backurs et al.~\cite{Backurs18} (arguably also following~\cite{Roditty13}) all the reductions feature layers $L_0, L_1, \ldots, L_{k-1}, L_k$, with only (forward) edges between two consecutive $L_i$. 
The vertices within the same layer share the same number of ``vector attributes'' and ``index attributes''.
The interplay between vector and index attributes in defining the vertices and edges is adjusted so that if there are no $k$~orthogonal vectors, then there are paths of ``optimal'' length $k$ between every pair in $L_0 \times L_k$, while if there is set~$X$ of $k$~orthogonal vectors, a pair $(x,y)$ in $L_0 \times L_k$ jointly encoding~$X$ is far apart (usually and ideally at distance $2k-1$).

We do not deviate too much from this strategy.
Our construction is inspired from and pushes one step forward the elegant reduction of Li~\cite{Li20}.
Here we rename $L_0, L_1, L_2, L_3, L_4$ by $\abc,\ab,\ady,\dc,$ $\dcb$, respectively.
Some pairs of vertices are too far apart on \fov NO-instances.
We thus add two ``gates'' $u$ and $v$ and link them with weighted arcs to the rest of the graph.
This puts many pairs of vertices at distance at most~4 regardless on whether the \fov instance is positive or negative.

Of course, we cannot do so for \emph{all} the pairs outside $L_0 \times L_k$.
For instance, we do not want the distance from every vertex in $\abc$ to every vertex in $\ady$ to be always at most~4.
Indeed, that would make the longest path from $\abc$ to $\dcb$ of length at most~6.
Our main novel contribution is to add a vertex set $\adx$ ($L'_2$) only linked to $\ady$ ($L_2$) in a way that gives enough flexibility to make the remaining pairs sufficiently close for \fov NO-instances, while not decreasing the distance from $x$ to $y$.

\paragraph*{Recent developments}

While the paper was under review, some exciting developments happened.
Independently, Wein and Dalirrooyfard~\cite{Wein20} and Li~\cite{Li20b} showed that, under the SETH, for every integer $k \geqslant 2$ and real $\delta > 0$, any $(\frac{2k-1}{k}-\delta)$-approximation for sparse unweigthed directed~\textsc{Diameter} requires time $n^{1+\frac{1}{k-1}-o(1)}$.

\begin{figure}[ht!]
  \centering
  \begin{tikzpicture}
    \def\s{8.5}
    \def\xb{1}
    \def\xe{2}
    \def\yb{1}
    \def\ye{2}
    \def\h{0.02}
    \draw[->] (\xb * \s, \yb * \s - \h * \s) --  (\xb * \s, \ye * \s + \h * \s) ;
    \draw[->] (\xb * \s - \h * \s, \yb * \s) --  (\xe * \s + \h * \s, \yb * \s) ;
    \foreach \i/\j in {1/1,1.1/{\textcolor{red!50!cyan}{$\frac{k+1}{k}$}},1.2/{6/5},1.25/{5/4},1.33/{4/3},1.5/{3/2},2/2}{
      \node at (\xb * \s - 0.05 * \s, \i * \s) {\j} ;
    }
    \foreach \i/\j in {1/1,1.5/{$\frac{3}{2}$},1.667/{$\frac{5}{3}$},1.75/{$\frac{7}{4}$},1.875/{\textcolor{red}{$\frac{2k+1}{k+1}$}},2/2}{
      \node at (\i * \s, \yb * \s - 0.05 * \s) {\j} ;
    }
    \node at (1.875 * \s, \yb * \s - 0.12 * \s) {\textcolor{cyan}{$\frac{2k-1}{k}$}} ;

    \node at (1.25 * \s, \yb * \s - 0.05 * \s) {approximation factor} ;
    \node at (\xb * \s - 0.05 * \s, 1.75 * \s) {\rotatebox{90}{runtime exponent}} ;
    \fill[red,opacity=0.3] (\xb * \s, \yb * \s) -- (1.5 * \s, \yb * \s) -- (1.5 * \s, \ye * \s) -- (\xb * \s, \ye * \s) -- cycle ;
    \node at (1.25 * \s, 1.5 * \s) {\cite[RV13]{Roditty13}} ;

    \fill[red,opacity=0.5] (1.5 * \s, \yb * \s) -- (1.667 * \s, \yb * \s) -- (1.667 * \s, 1.5 * \s) -- (1.5 * \s, 1.5 * \s) -- cycle ;
    \node at (1.585 * \s, 1.25 * \s) {\cite[Li20]{Li20}} ;

    \fill[red,opacity=0.7] (\xe * \s, \yb * \s) -- (1.889 * \s, 1.111 * \s) -- (1.889 * \s, 1.125 * \s) -- (1.875 * \s, 1.125 * \s) -- (1.875 * \s, 1.1428 * \s) -- (1.8571 * \s, 1.1428 * \s) -- (1.8571 * \s, 1.1667 * \s) -- (1.833 * \s, 1.1667 * \s) -- (1.833 * \s, 1.2 * \s) -- (1.8 * \s, 1.2 * \s) -- (1.8 * \s, 1.25 * \s) -- (1.75 * \s, 1.25 * \s) -- (1.75 * \s, 1.33 * \s) -- (1.667 * \s, 1.33 * \s) -- (1.667 * \s, \yb * \s) -- cycle ;

    \node at (1.78 * \s, 1.1 * \s) {\cite[Li20]{Li20b}} ;
     \node at (1.78 * \s, 1.04 * \s) {\cite[DW20]{Wein20}} ;
    \fill[blue,opacity=0.25] (1.5 * \s, 1.5 * \s) -- (1.5 * \s, \ye * \s) -- (\xe * \s, \ye * \s) -- (\xe * \s, 1.5 * \s) -- cycle ;
    \node at (1.75 * \s, 1.8 * \s) {\cite[RV13]{Roditty13}} ;
    \node at (1.75 * \s, 1.7 * \s) {\cite[CLRSTV14]{Chechik14}} ;

     \begin{scope}
       \pgfsetfillpattern{dots}{cyan}
     \fill (1.889 * \s, 1.111 * \s) -- (1.889 * \s, 1.125 * \s) -- (1.875 * \s, 1.125 * \s) -- (1.875 * \s, 1.1428 * \s) -- (1.8571 * \s, 1.1428 * \s) -- (1.8571 * \s, 1.1667 * \s) -- (1.833 * \s, 1.1667 * \s) -- (1.833 * \s, 1.2 * \s) -- (1.8 * \s, 1.2 * \s) -- (1.8 * \s, 1.25 * \s) -- (1.75 * \s, 1.25 * \s) -- (1.75 * \s, 1.33 * \s) -- (1.667 * \s, 1.33 * \s) -- (1.667 * \s, 1.5 * \s) -- (\xe * \s, 1.5 * \s) -- (\xe * \s, \yb * \s) -- cycle;
     \end{scope}
    \draw (\xe * \s, \yb * \s) -- (1.889 * \s, 1.111 * \s) -- (1.889 * \s, 1.125 * \s) -- (1.875 * \s, 1.125 * \s) -- (1.875 * \s, 1.1428 * \s) -- (1.8571 * \s, 1.1428 * \s) -- (1.8571 * \s, 1.1667 * \s) -- (1.833 * \s, 1.1667 * \s) -- (1.833 * \s, 1.2 * \s) -- (1.8 * \s, 1.2 * \s) -- (1.8 * \s, 1.25 * \s) -- (1.75 * \s, 1.25 * \s) -- (1.75 * \s, 1.33 * \s) -- (1.667 * \s, 1.33 * \s) -- (1.667 * \s, 1.5 * \s) -- (1.5 * \s, 1.5 * \s) -- (1.5 * \s, \ye * \s) -- (\xb * \s, \ye * \s) ;

    \node at (1.875 * \s, 1.32 * \s) {\cite[Li20]{Li20b}} ;
    \foreach \i/\j/\c in {1/2/blue,1.5/2/red,1.5/1.5/blue,1.667/1.5/red,1.75/1.33/red,2/1/blue, 1.667/1.33/cyan,1.75/1.25/cyan, 1.8/1.25/red, 1.8/1.2/cyan, 1.833/1.2/red,1.833/1.1667/cyan,1.8571/1.1667/red,1.8571/1.1428/cyan,1.875/1.1428/red,1.875/1.125/cyan,1.889/1.125/red,1.889/1.111/cyan}{
      \draw[\c] (\i * \s, \j * \s) circle [radius=1.2pt] ;
      \fill[\c,opacity=0.8] (\i * \s, \j * \s) circle [radius=1.2pt] ;
    }
  \end{tikzpicture}
  \caption{The new results for sparse unweigthed directed~\textsc{Diameter}.
  The blue region is feasible, as witnessed by an algorithm at the bottom-left corner (blue dot).
  The red regions would refute the SETH, as witnessed by reductions at top-right corners (red dots).
  A red dot in the interior of the dotted cyan region would refute the NSETH.
  The lower bounds in~\cite{Roditty13,Li20} even hold for the sparse unweighted undirected~\textsc{Diameter}.
  }
  \label{fig:new-results}
\end{figure}
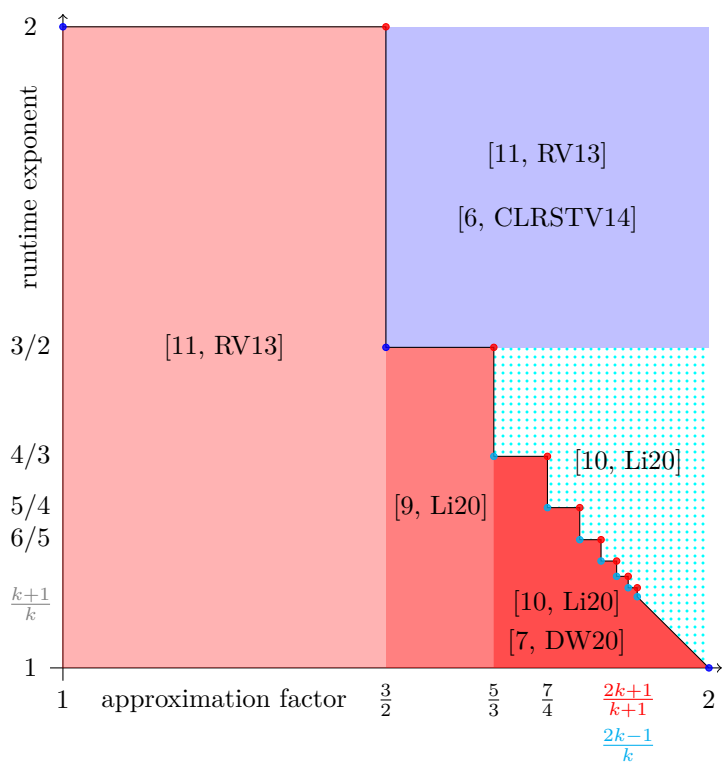

In the same paper, Li gives a piece of evidence that this could be as far as the hardness of unweigthed directed~\textsc{Diameter} goes.
This evidence is based on the NSETH (for Nondeterministic SETH), a strengthening of SETH introduced by Carmosino et al.~\cite{Carmosino16}.
NSETH asserts that for every $\varepsilon > 0$, there is an integer $k$ such that the \textsc{$k$-Taut} problem cannot be solved in \emph{non-deterministic} time $(2-\varepsilon)^n$, where \textsc{$k$-Taut} asks, given a $k$-DNF formula whether every truth assignment satisfies it. 
Li shows, for all four variants of~\textsc{Diameter} \emph{but} the weighted directed one that improving on any of these (deterministic) SETH lower bounds would refute the NSETH (see dotted cyan region in~\cref{fig:new-results}). 

The construction of Dalirrooyfard and Wein~\cite{Wein20} works for $k \geqslant 5$.
It makes a more intricate use of ``parallel layers'' (such as $\adx$).
For $k=2$ and $3$, the authors cite the existing lower bounds for unweighted undirected~\textsc{Diameter}, and for $k=4$ they tune the construction presented in this paper, in order to remove the edge weights.
Namely, there is a simpler and weight-free way of connecting the ``gates'' $u$, $v$ to the rest of the graph.
The construction of Li~\cite{Li20b} effectively combines \emph{index-changing} ``back'' edges (which re-implements and extends the \emph{skew edges} of this paper) with the usual \emph{vector-changing} ``forward'' edges.

These generalizations crucially rely on edge orientations, while in the particular case of $k=4$, our construction seems to only accidentally require arcs.
In a subsequent work~\cite{Bonnet21}, we show how to obtain the lower bound for $k=4$ in the most constrained case of unweigthed \emph{undirected}~\textsc{Diameter}.
This makes a non-trivial use of additional sets of vertices without ``vector fields''.

\paragraph*{Preliminaries}

We use the standard graph-theoretic notations.
If $G$ is a graph, $V(G)$ denotes its vertex set.
If $S \subseteq V(G)$, $G[S]$ denotes the subgraph of $G$ induced by $S$, and $G - S$ is a short-hand for $G[V(G) \setminus S]$.
For $u, v \in V(G)$, $\dist_G(u,v)$ denotes the distance from $u$ to $v$ in $G$, that is the length of a shortest path from $u$ to $v$, or equivalently, the minimum sum of weights on the edges on a path from $u$ to $v$.
Note that, in a directed graph, $\dist_G(u,v)$ and $\dist_G(v,u)$ may well be different values.
We drop the subscript, if the graph $G$ is clear from the context.
We denote by $\diam(G)$ the diameter of $G$, that is, $\max_{u,v \in V(G)} \dist_G(u,v)$.
Note that both the pairs $(u,v)$ and $(v,u)$ are considered in this maximum.
If $\ell$ is positive integer, $[\ell]$ denotes the set $\{1,2,\ldots,\ell\}$. 

\section{Reduction from 4-Orthogonal Vectors to 4 vs 7 Diameter}

For every fixed positive integer $k$, the \lkov (\kov for short) problem is as follows.
It asks, given a set $S$ of 0,1-vectors in $\{0,1\}^\ell$, if there are $k$ vectors $v_1, \ldots, v_k \in S$ such that for every $i \in [\ell]$, $\Pi_{h \in [k]} v_h[i] = 0$ or equivalently that $v_1[i] = v_2[i] = \cdots = v_k[i] = 1$ does not hold.
Williams~\cite{Williams05} showed that, assuming the SETH, \kov requires $N^{k-o(1)}$ time with $N := |S|$.
Here we will leverage this lower bound for $k=4$.
This is a usual opening step: for example, Roditty and Vassilevska Williams~\cite{Roditty13} uses this lower bound for $k = 2$, and Li~\cite{Li20} uses it for $k=3$.

From any set $S$ of $N$ vectors in $\{0,1\}^\ell$, we build a directed weighted graph $G := \rho(S)$ (without negatively-weighted arcs) with $O(N^3+N^2\ell^3)$ vertices and $O(N^3\ell^3+N^2\ell^6)$ arcs such that if $S$ admits an orthogonal quadruple then the diameter of $G$ is (at least) 7, whereas if $S$ has no orthogonal quadruple then the diameter of $G$ is (at most) 4.
There is a large enough constant $c$ such that \fov requires $N^{4-o(1)}$ time, unless the SETH fails, even when $\ell = c \lceil \log N \rceil$~\cite{Williams05}.
In that case, the graph $G$ has $O(N^3)$ vertices and $\tilde{O}(N^3)$ edges.
Hence any algorithm approximating sparse, weighted, directed \textsc{Diameter} within ratio better than $7/4$ in time $n^{4/3 - \delta}$, with $\delta > 0$, would refute the SETH.

\subsection{Constant part}\label{sec:constant}

\begin{figure}
  \centering
  \begin{tikzpicture}
    \def\h{3}
    \def\v{2}
    \foreach \i/\j/\l/\ll in {0/0/ABC/abc, 0/1/AB/ab, 1/1/$\text{AD}_X$/adx, 1/2/$\text{AD}_Y$/ady, 2/1/DC/dc, 2/0/DCB/dcb}{
      \node[draw, rounded corners] (\ll) at (\h * \i,\v * \j) {\l} ;
    }
    \foreach \i/\j/\l in {0.7/-0.6/u,1.3/-0.6/v}{
      \node[draw, circle] (\l) at (\h * \i,\v * \j) {\l} ;
    }

    \foreach \i/\j/\w/\posi/\mdw/\b in {abc.south/u/4/below/0.5/0, u/abc.east/0/above/0.5/0, v/dcb.west/4/above/0.5/0, dcb.south/v/0/below/0.5/0, u/ab/0/below/0.7/0, dc/v/0/below/0.3/0, ab/u/3/above/0.5/15, v/dc/3/above/0.5/15}{
      \draw[-stealth] (\i) to [bend left=\b] node[style={pos=\mdw}, \posi] {\w} (\j) ;
    }
    \foreach \i/\j/\w/\posi/\mdw/\b in {u/v/2/below/0.5/0, u/adx/1/left/0.75/0, v/adx/1/right/0.75/0, u/ady/2/left/0.7/10, v/ady/2/right/0.7/-10, ab.east/v/2/above/0.25/10, dc.west/u/2/above/0.25/-10}{
      \draw (\i) to [bend left=\b] node[style={pos=\mdw}, \posi] {\w} (\j) ;
    }
    \foreach \i/\j in {abc/ab, ab/ady, ady/adx, ady/dc, dc/dcb}{
      \draw[very thick, red] (\i) -- (\j) ;
    }
  \end{tikzpicture}
  \caption{The part of the reduction \emph{not} depending on the \fov instance.
    The edges represented without arrow are double-arcs with the indicated weight.
    The black arc between, say, \abc and $u$, symbolizes that every vertex of \abc is linked by an arc of weight 4 to vertex $u$.
    Thick red edges represent \emph{some} double-arcs of weight 1.
    Not every double-arc (or edge) is present between two sets linked by a red edge.
    This will be specified in the rest of the construction.}
  \label{fig:constant}
\end{figure}
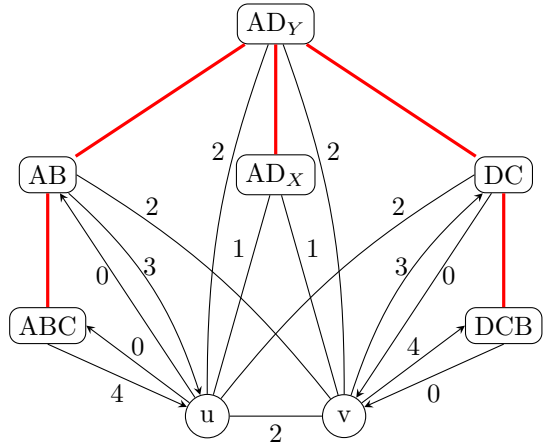

We start by describing the part of the construction which does \emph{not} depend on the \fov instance.
Its purpose is to make many pairs of vertices at distance at most~4 regardless on whether the \fov instance is positive or negative.
The vertex set of the eventually-built graph $G$ consists of two special vertices $u$ and $v$, and six (disjoint) sets \abc, \ab, $\adx$, $\ady$, \dc, and \dcb.
Vertices $u$ and $v$ are unconditionally linked to these sets (and to each other) by weighted arcs as specified in~\cref{fig:constant}.
As we wrote in the introduction, there is a simpler way, that does not require edge weights, of realizing the constant part (see \cite[Section 6]{Wein20}).
We keep our construction for the sake of consistency but invite the reader to have a look at \cite[Figure 9]{Wein20}.

In this figure, a black arc between a vertex $x \in \{u,v\}$ and a set $Z \in \{\abc, \ab, \adx, \ady,$ $\dc, \dcb\}$ (or vice versa) indicates that $x$ is linked to \emph{every vertex} of $Z$ by such an arc. 
Note that edges represented without arrow are double-arcs.
Double-arcs will sometimes simply be called \emph{edges}.
The only arcs \emph{not} incident to $\{u,v\}$ are edges (double-arcs) of weight 1.
These edges are only present between \abc and \ab, \ab and $\ady$, $\adx$ and $\ady$, $\ady$ and \dc, and finally \dc and \dcb.
We will describe them later.
At this point, one just needs to know that every vertex in \abc (resp.~\dcb) has at least one neighbor in \ab (resp.~\dc), and that these edges have weight~1.

We check that many pairs of vertices are at distance at most 4 (even without the knowledge of the edges symbolized in red).
For the sake of conciseness, when we write, say, ``$\mathbf{u \leftrightarrow}~\textbf{\dcb}$'', we intend to provide paths from $u$ to every vertex in $\dcb$, and from every vertex in $\dcb$ to~$u$.
Similarly, the paragraph ``$\textbf{\dcb}~\mathbf{\rightarrow}~\textbf{\abc}$'' gives a path (of length~2) from every vertex of \dcb to every vertex of \abc. 

\medskip

$\mathbf{u \leftrightarrow v}.$ There is an edge of weight 2 between $u$ and $v$.

\medskip

$\mathbf{u \leftrightarrow Z} \in \{\abc,\ab,\adx,\ady,\dc\}.$ There are double-arcs with weight at most 4 between these pairs.

\medskip

$\mathbf{u \leftrightarrow}~\textbf{\dcb}.$ There is a path of double-arcs of total weight 3: From a vertex of \dcb, take any (weight-1) edge to \dc, followed by the weight-2 edge to $u$.
Recall that every vertex in \dcb will have at least one neighbor in \dc (via a weight-1 edge).

\medskip

The next two cases are symmetric.

\medskip

$\mathbf{v \leftrightarrow Z} \in \{\ab,\adx,\ady,\dc,\dcb\}.$ There are double-arcs with weight at most 4 between these pairs.

\medskip

$\mathbf{v \leftrightarrow}~\textbf{\abc}.$ There is a path of double-arcs of total weight 3: From a vertex of \abc, take any (weight-1) edge to \ab, followed by the weight-2 edge to $v$.
Recall that every vertex in \abc will have at least one neighbor in \ab (via a weight-1 edge).

\medskip

So far, we have seen that for each $x \in \{u,v\}$ and $y \in V(G)$, $\dist(x,y) \leqslant 4$ and $\dist(y,x) \leqslant 4$.

\medskip

\textbf{For each}~$\mathbf{Z \in \{\abc,\ab,\adx,\ady,\dc,\dcb\}}$\textbf{,}~$\mathbf{Z \leftrightarrow Z}.$
Each of these six sets $Z$ has a double-arc to $u$ or to $v$ (or both) whose sum of weights is at most~4.

\medskip

\textbf{AD}$\mathbf{_X \rightarrow Z} \in \{\abc,\ab,\ady,\dc\}.$
These pairs are at distance at most 3.
There is an edge of weight 1 from every vertex of $\adx$ to $u$, and an arc of weight at most 2 from $u$ to every vertex of $Z \in \{\abc,\ab,\ady,\dc\}$.

\medskip

\textbf{AD}$\mathbf{_X \rightarrow}~\textbf{\dcb}.$
There is a path of length 4, via $u$ and \dc.
Again recall that every vertex of \dcb has at least one neighbor in \dc (via a double-arc of weight 1).

\medskip

The next two cases are symmetric.

\medskip

\textbf{AD}$\mathbf{_X \leftarrow Z} \in \{\ab,\ady,\dc,\dcb\}.$
These pairs are at distance at most 3.
There is an arc of weight at most 2 from every vertex of $Z \in \{\ab,\ady,\dc,\dcb\}$ to $v$, and an edge of weight 1 from $v$ to every vertex of $\adx$.

\medskip

\textbf{AD}$\mathbf{_X \leftarrow}~\textbf{\abc}.$
There is path of length 4, via \ab and $v$.

\medskip

We have now established that for every $x \in \adx$ and $y \in V(G)$, $\dist(x,y) \leqslant 4$ and $\dist(y,x) \leqslant 4$.

\medskip

\textbf{AD}$\mathbf{_Y \leftrightarrow}~\textbf{\ab}.$
There is a path of two double-arcs of weight 2, via $v$. 

\medskip

\textbf{AD}$\mathbf{_Y \leftrightarrow}~\textbf{\dc}.$
There is a path of two double-arcs of weight 2, via $u$.

\medskip

\textbf{AD}$\mathbf{_Y \rightarrow}~\textbf{\abc}.$
There is a path of length 2 via $u$.

\medskip

\textbf{AD}$\mathbf{_Y \leftarrow}~\textbf{\dcb}.$
There is a path of length 2 via $v$.

\medskip

$\textbf{\abc}~\mathbf{\leftrightarrow}~\textbf{\ab}.$
Via $u$, there is a path of length 4 from every vertex of \abc to every vertex of \ab, and a path of length 3 from every vertex of \ab to every vertex of \abc.

\medskip

$\textbf{\dcb}~\mathbf{\leftrightarrow}~\textbf{\dc}.$
Via $v$, there is a path of length 3 from every vertex of \dcb to every vertex of \dc, and a path of length 4 from every vertex of \dc to every vertex of \dcb.

\medskip

$\textbf{\dcb}~\mathbf{\rightarrow}~\textbf{\abc}.$
There is a path of length 2 via $v$ and $u$.

\medskip

$\textbf{\dcb}~\mathbf{\rightarrow}~\textbf{\ab}.$
There is a path of length 2 via $v$.

\medskip

$\textbf{\dc}~\mathbf{\rightarrow Z} \in \{\abc,\ab\}.$ 
There is a path of length 2 via $u$.

\medskip

To summarize, we have obtained that for every pair $x, y \in V(G)$, $\dist(x,y) > 4$ implies that $(x,y) \in P$ for some $P \in \{\abc \times \ady, \abc \times \dc, \abc \times \dcb, \ab \times \dc, \ab \times \dcb, \ady \times \dcb\}$.

\subsection{Variable part}\label{sec:variable}

We think of the vector set $S$ as having four copies $A, B, C, D$ with $S = A = B = C = D$.
Equivalently, one looks for $a \in A$, $b \in B$, $c \in C$, and $d \in D$ such that $a, b, c, d$ are orthogonal.

\paragraph*{Vertex set}

We first describe the vertex set $V(G) \setminus \{u,v\}$.
\begin{itemize}
\item For every $(a,b,c) \in A \times B \times C$, we add vertex $(a,b,c)_{\abc}$ to $\abc$.
\item Similarly for every $(d,c,b) \in D \times C \times B$, we add vertex $(d,c,b)_{\dcb}$ to $\dcb$.
\item For every $(a,b) \in A \times B$ and every triple $i, j, k \in [\ell]$ such that $a[i] = a[j] = a[k] = 1$ and $b$ takes value 1 on at least two indices of $\{i,j,k\}$, we add vertex $(a,b,i,j,k)_{\ab}$ to $\ab$.
\item The set of vertices $\dc$ is defined analogously with $D$ and $C$ playing the roles of $A$ and $B$ (recall that actually $A=B=C=D$).
\item For every $(a,d) \in A \times D$ and every triple of indices $i, j, k \in [\ell]$ such that $a[i] = a[j] = a[k] = 1$ and $d[i] = d[j] = d[k]=1$, we add vertex $(a,d,i,j,k)_{\ady}$ to $\ady$.
\item For every $(a,d) \in A \times D$ and every $i, j, k \in [\ell]$ such that at most one of $a[i], a[j], a[k], d[i],$ $d[j], d[k]$ is equal to 0, we add vertex $(a,d,i,j,k)_{\adx}$ to $\adx$.
\end{itemize}

As observed in~\cite{Wein20}, the definition of $\adx$ could be simpler.
We keep it as is, again for the sake of consistency.

\paragraph*{Edge Set}

We now describe the edge set on $V(G) \setminus \{u,v\}$.
All these edges are double-arcs (we do not need to orient them) of weight 1 (we do not need to put weights). 

\begin{itemize}
\item We add an edge (double-arc) of weight 1 between every pair $(a,b,c)_{\abc}$ and $(a,b,i,j,k)_{\ab}$ if $c$ takes value 1 on at least one index of $\{i,j,k\}$ where $b$ also takes value 1.
In our construction, the existence of an edge is implicitly conditional to the existence of both of its endpoints.
The edge exists only if $(a,b,i,j,k)_{\ab}$ is indeed a vertex of $\ab$.
  The edges between $\dcb$ and $\dc$ are defined similarly.
\item
We add every edge between $(a,b,i,j,k)_\ab$ and $(a,b,i',j',k')_\ab$, with $a \in A$, $b \in B$, and $i,j,k,i',j',k' \in [\ell]$.
Similarly we add every edge between $(d,c,i,j,k)_\dc$ and $(d,c,i',j',k')_\dc$, with $d \in D$, $c \in C$, and $i,j,k,i',j',k' \in [\ell]$.
We call these edges \emph{index-switching}.
$\ab$ and $\dc$ are the only two sets among $\{\abc,\ab,\adx,\ady,\dc,\dcb\}$ which are \emph{not} independent sets.
\item
We link every pair $(a,b,i,j,k)_\ab$ and $(a,d,i,j,k)_{\ady}$ by an edge, as well as every pair $(a,d,i,j,k)_{\ady}$ and $(d,c,i,j,k)_\dc$.
\item
  We add an edge between every pair $(a,d,i,j,k)_{\adx}$ and $(a,d',i,j,k)_{\ady}$ (such that $a \in A$, $d \neq d' \in D$, and $i, j, k \in [\ell]$), and $(a,d,i,j,k)_{\adx}$ and $(a',d,i,j,k)_{\ady}$ (such that $a \neq a' \in A$, $d \in D$, and $i, j, k \in [\ell]$).
\item
Finally we add an edge between every pair $(a,d,i,j,k)_{\adx}$ and $(a,d,i',j',k')_{\ady}$ with $a \in A$, $d \in D$, and $i, j, k, i', j', k' \in [\ell]$.
We call this type of edge \emph{skew}.
Skew edges are the only way to change the indices while also moving from one set to another (it is not internal to \ab or to \dc).
\end{itemize}

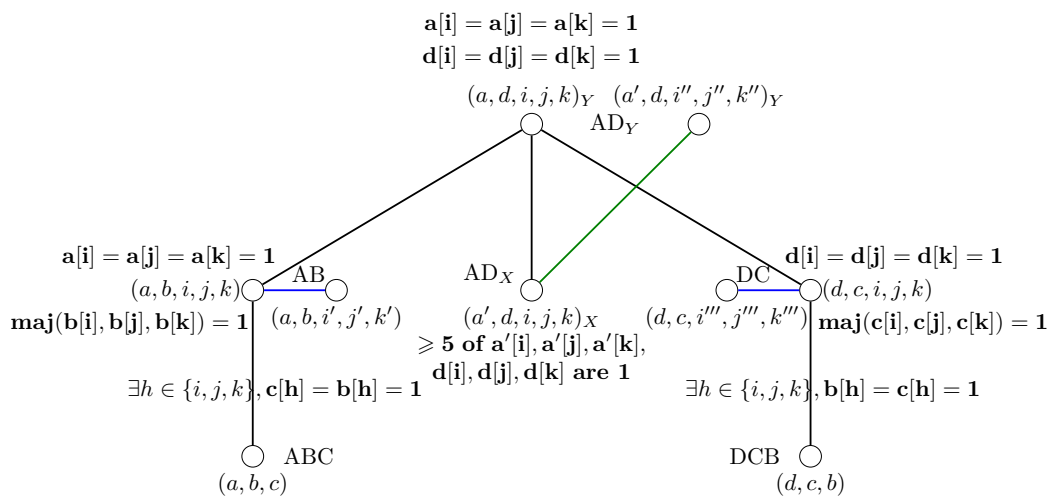
\begin{figure}
  \centering
  \resizebox{400pt}{!}{
  \begin{tikzpicture}[scale=0.85]
    \def\h{5}
    \def\v{3}
    \foreach \i/\j/\l/\ll/\x/\y in {0/0/{(a,b,c)}/abc/0/-0.5, 0/1/{(a,b,i,j,k)}/abijk/-1.2/0, 0.3/1/{(a,b,i',j',k')}/abijk2/0/-0.5, 1/1/{(a',d,i,j,k)_X}/adijkx/0/-0.5, 1/2/{(a,d,i,j,k)_Y}/adijky/0/0.5, 1.6/2/{(a',d,i'',j'',k'')_Y}/adijky2/0/0.5, 2/1/{(d,c,i,j,k)}/dcijk/1.2/0, 1.7/1/{(d,c,i''',j''',k''')}/dcijk2/0/-0.5, 2/0/{(d,c,b)}/dcb/0/-0.5}{
      \node[draw, circle] (\ll) at (\h * \i,\v * \j) {} ;
      \node (t\ll) at (\h * \i + \x,\v * \j + \y) {$\l$} ;
    }

    \foreach \i/\j/\c in {abc/abijk/black, abijk/adijky/black, adijkx/adijky/black, adijky/dcijk/black, dcijk/dcb/black, abijk/abijk2/blue, dcijk/dcijk2/blue, adijkx/adijky2/green!50!black}{
      \draw[thick, color=\c] (\i) -- (\j) ;
    }

    \node (ab1) at (- 0.3 * \h, 1.2 * \v) {$\mathbf{a[i]=a[j]=a[k]=1}$} ;
    \node (ab2) at (- 0.44 * \h, 0.8 * \v) {\textbf{maj}$\mathbf{(b[i],b[j],b[k])=1}$} ;
    \node (cd1) at (2.3 * \h, 1.2 * \v) {$\mathbf{d[i]=d[j]=d[k]=1}$} ;
    \node (cd2) at (2.44 * \h, 0.8 * \v) {\textbf{maj}$\mathbf{(c[i],c[j],c[k])=1}$} ;

    \node at (0.08 * \h, 0.4 * \v) {$\exists h \in \{i,j,k\}, \mathbf{c[h]=b[h]=1}$} ;
    \node at (2.08 * \h, 0.4 * \v) {$\exists h \in \{i,j,k\}, \mathbf{b[h]=c[h]=1}$} ;

    \node (ady1) at (1 * \h, 2.4 * \v) {$\mathbf{d[i]=d[j]=d[k]=1}$} ;
    \node (ady1) at (1 * \h, 2.6 * \v) {$\mathbf{a[i]=a[j]=a[k]=1}$} ;

    \node (adx1) at (1 * \h, 0.67 * \v) {$\mathbf{\geqslant 5~\textbf{of}~a'[i], a'[j], a'[k],}$} ;
    \node (adx1) at (1 * \h, 0.5 * \v) {$\mathbf{d[i], d[j], d[k]}$~\textbf{are 1}} ;

    \node at (0.2 * \h,0) {\abc} ;
    \node at (1.8 * \h,0) {\dcb} ;
    \node at (0.2 * \h,1.1 * \v) {\ab} ;
    \node at (1.795 * \h,1.1 * \v) {\dc} ;
    \node at (0.85 * \h,1.1 * \v) {$\adx$} ;
    \node at (1.3 * \h,2 * \v) {$\ady$} ;
  \end{tikzpicture}
  }
  \caption{The rules for the existence of vertices and edges in $G - \{u,v\}$.
    We removed the subscripts in the vertex labels as discussed in the second-to-last paragraph of~\cref{sec:variable}.
    Conditions to the existence of a vertex appear in bold next to the vertex.
    Conditions to the existence of an edge appear in bold along the edge.
    All edges (double-arcs) have weight 1, so we omit their weight.
    Regular edges are represented in black.
    Index-switching edges are represented in blue (they are only present in $\ab$ and $\dc$).
    Skew edges are represented in green (they are only present between $\adx$ and $\ady$.
    Note that the regular edge $(a,d,i,j,k)_Y(a',d,i,j,k)_X$ could also be of the form $(a,d,i,j,k)_Y(a,d',i,j,k)_X$.}
  \label{fig:variable}
\end{figure}

All the edges defined in this section that are not index-switching or skew are called \emph{regular}.
This ends the construction of $G = \rho(S)$.
See~\cref{fig:variable} for an illustration.

\paragraph*{Simplified vertex notations, vector fields, and index fields}

Henceforth we will drop the indices in the vertex labels.
The set a vertex belongs to will be implicit by the choice of the variable labels.
For instance $(a,b',i,j,k)$ is in $\ab$, and $(d',c,b')$ is in $\dcb$.
This does not allow us to distinguish vertices of $\adx$ and $\ady$.
We will denote by $(a,d,i,j,k)_Y = (a,d,i,j,k)_{\ady}$ a vertex in $\ady$ and by $(a,d,i,j,k)_X$ the ``same'' vertex in $\adx$.
Note that it is possible that $(a,d,i,j,k)_X$ exists but not $(a,d,i,j,k)_Y$, if exactly five of $a[i], a[j], a[k], d[i], d[j], d[k]$ are equal to 1.

We call \emph{vector fields} the first three coordinates of every vertex in $\abc \cup \dcb$, and the first two coordinates of every vertex in $\ab \cup \adx \cup \ady \cup \dc$.
We call \emph{index fields} the last three coordinates of every vertex in $\ab \cup \adx \cup \ady \cup \dc$.
We can assume that the \fov instance does not have an orthogonal \emph{triple} (this can be checked in time $\tilde{O}(N^3)$).
Thus every vertex $(a,b,c) \in \abc$ (resp.~$(d,c,b) \in \dcb$) indeed has at least one neighbor in $\ab$ (resp.~$\dc$), namely $(a,b,i,i,i)$ (resp.~$(d,c,i,i,i)$) where $i \in [\ell]$ is such that $a[i] = b[i] = c[i] = 1$ (resp.~$d[i] = c[i] = b[i] = 1$).

\paragraph*{Vertex and edge count}

Before tackling the correctness of the reduction, we check that $G$ has $O(N^3)$ vertices and $\Tilde{O}(N^3)$ arcs.
The number of vertices of $G$ is bounded by $2+2 \cdot N^3+4 \cdot N^2 \ell^3 = O(N^3)$ since $\ell = O(\log N)$.
The number of arcs of $G$ is bounded by $4 \cdot O(N^3) + 4 \cdot N^2 \ell^6 + 10 \cdot N^3 \ell^3 + 2 \cdot N^2 \ell^6 = \Tilde{O}(N^3)$, where the first term accounts for the arcs incident with $\{u,v\}$, the second for the index-switching arcs, the third for the regular arcs, and the fourth for the skew arcs.

\subsection{No orthogonal quadruple implies diameter at most 4}

We exhibit in this section short paths (of length at most 4) between every pair of vertices of $G$.
For that we extensively use that, as there is no orthogonal quadruple in $S$, for every $u,v,w,x \in S$, $\ind(u,v,w,x) := \min\{i \in [\ell] $ $|$ $u[i]=v[i]=w[i]=x[i]=1\}$ is a well-defined index in $[\ell]$.
We only take the minimum index to have a deterministic notation.
There will not be anything particular with the minimum, and any index of the non-empty $\{i \in [\ell] $ $|$ $u[i]=v[i]=w[i]=x[i]=1\}$ would work as well.
We will also use $\ind(u,v,w)$ as a short-hand for $\ind(u,v,w,w)$.

In~\cref{sec:constant}, we have reduced the task of showing that $\diam(G) \leqslant 4$ to considering only six pairs of sets.

\medskip

$\textbf{\abc}~\mathbf{\rightarrow}~\textbf{AD}\mathbf{_Y}.$
Let $(a,b,c)$ and $(a',d,i',j',k')_Y$ be two vertices in $\abc$ and $\ady$ respectively.
We define the indices $i := \ind(a,b,c,d)$, $j := \ind(a,a',b,d)$, and $k := \ind(a,a',d)$.
Then, $(a,b,c) \rightarrow (a,b,i,j,k) \rightarrow (a,d,i,j,k)_Y \rightarrow (a',d,i,j,k)_X \rightarrow (a',d,i',j',k')_Y$ is a path of length~4 in~$G$.

We first justify the existence of the inner vertices of this path (i.e., all but the endpoints).
Indeed the endpoints exist by assumption.
Vertex $(a,b,i,j,k) \in \ab$ is present in $G$ since $a[i] = a[j] = a[k] = 1$, and $b[i] = b[j] = 1$.
Vertex $(a,d,i,j,k)_Y \in \ady$ exists since $a[i] = a[j] = a[k] = 1 = d[i] = d[j] = d[k]$.
Finally $(a',d,i,j,k)_X \in \adx$ is indeed a vertex of $G$ since $a'[j] = a'[k] = 1$ and $d[i] = d[j] = d[k] = 1$.
Recall that in $\adx$ (contrary to $\ady$) it is fine if at most one of the six values obtained by evaluating one of the two vectors at one of the three indices is 0.

We now justify the existence of the edges.
The arc $(a,b,c) \rightarrow (a,b,i,j,k)$ exists since $c[i] = b[i] = 1$ (and both its endpoints exist).
The arcs $(a,b,i,j,k) \rightarrow (a,d,i,j,k)_Y$ and $(a,d,i,j,k)_Y \rightarrow (a',d,i,j,k)_X$ are regular edges of $G$: one vector field and the three index fields remain unchanged.
Finally the arc $(a',d,i,j,k)_X \rightarrow (a',d,i',j',k')_Y$ is a skew edge of~$G$: it is between $\adx$ and $\ady$, and both vector fields remain the same (while the indices are allowed to change). 

\medskip

$\textbf{\abc}~\mathbf{\rightarrow}~\textbf{\dc}.$
Let $(a,b,c)$ and $(d,c',i',j',k')$ be two vertices in $\abc$ and $\dc$ respectively.
We define the indices $i := \ind(a,b,c,d)$, $j := \ind(a,b,c',d)$, and $k := \ind(a,c',d)$.
Then, $(a,b,c) \rightarrow (a,b,i,j,k) \rightarrow (a,d,i,j,k)_Y \rightarrow (d,c',i,j,k) \rightarrow (d,c',i',j',k')$ is a path of length~4 in~$G$.

As in the previous case, the existence of vertices $(a,b,i,j,k) \in \ab$, $(a,d,i,j,k)_Y \in \ady$, $(d,c',i,j,k) \in \dc$ is ensured by the fact that $a[i] = a[j] = a[k] = 1 = d[i] = d[j] = d[k]$ and $b[i] = b[j] = 1 = c'[j] = c'[k]$.
The arc $(a,b,c) \rightarrow (a,b,i,j,k)$ exists since $c[i] = b[i] = 1$, and the arcs $(a,b,i,j,k) \rightarrow (a,d,i,j,k)_Y \rightarrow (d,c',i,j,k)$ are two (existing) regular arcs.
Finally $(d,c',i,j,k) \rightarrow (d,c',i',j',k')$ is an index-switching arc internal to $\dc$ (note that the two vector fields remain the same, as they should). 

\medskip

$\textbf{\abc}~\mathbf{\rightarrow}~\textbf{\dcb}.$
Let $(a,b,c)$ and $(d,c',b')$ be two vertices in $\abc$ and $\dcb$ respectively.
We define the indices $i := \ind(a,b,c,d)$, $j := \ind(a,b,c',d)$, and $k := \ind(a,b',c',d)$.
Then, $(a,b,c) \rightarrow (a,b,i,j,k) \rightarrow (a,d,i,j,k)_Y \rightarrow (d,c',i,j,k) \rightarrow (d,c',b')$ is a path of length~4 in~$G$.

The vertices $(a,b,i,j,k) \in \ab$, $(a,d,i,j,k)_Y \in \ady$, $(d,c',i,j,k) \in \dc$ exist since $a[i] = a[j] = a[k] = 1 = d[i] = d[j] = d[k]$ and $b[i] = b[j] = 1 = c'[j] = c'[k]$.
The arc $(a,b,c) \rightarrow (a,b,i,j,k)$ is in $G$ since $c[i] = b[i] = 1$.
The arcs $(a,b,i,j,k) \rightarrow (a,d,i,j,k)_Y \rightarrow (d,c',i,j,k)$ are two regular arcs in $G$.
The arc $(d,c',i,j,k) \rightarrow (d,c',b')$ exists since $b'[k] = c'[k] = 1$.

\medskip

$\textbf{\ab}~\mathbf{\rightarrow}~\textbf{\dc}.$
Let $(a,b,i',j',k')$ and $(d,c,i'',j'',k'')$ be two vertices in $\ab$ and $\dc$ respectively.
We define the index $i := \ind(a,b,c,d)$.
Then, $(a,b,i',j',k') \rightarrow (a,b,i,i,i) \rightarrow (a,d,i,i,i)_Y \rightarrow (d,c,i,i,i) \rightarrow (d,c,i'',j'',k'')$ is a path of length~4 in~$G$.

Vertices $(a,b,i,i,i) \in \ab$, $(a,d,i,i,i)_Y \in \ady$, $(d,c,i,i,i) \in \dc$ exist since $a[i] = b[i] = c[i] = d[i] = 1$.
The arc $(a,b,i',j',k') \rightarrow (a,b,i,i,i)$ is a legal index-switching arc, internal to $\ab$.
The arcs $(a,b,i,i,i) \rightarrow (a,d,i,i,i)_Y \rightarrow (d,c,i,i,i)$ are two regular arcs in $G$.
Finally the arc $(d,c,i,i,i) \rightarrow (d,c,i'',j'',k'')$ is an index-switching arc, internal to $\dc$.

\medskip

The next two cases are symmetric to $\textbf{\abc}~\mathbf{\rightarrow}~\textbf{\dc}$ and $\textbf{\abc}~\mathbf{\rightarrow}~\textbf{AD}\mathbf{_Y}$, respectively.
We spell them out since it is not much longer that making the symmetry explicit.

\medskip

$\textbf{\ab}~\mathbf{\rightarrow}~\textbf{\dcb}.$
Let $(a,b,i',j',k')$ and $(d,c,b')$ be two vertices in $\ab$ and $\dcb$ respectively.
We define the indices $i := \ind(a,b,c,d)$, $j := \ind(a,b,d)$, and $k := \ind(a,b',c,d)$.
Then, $(a,b,i',j',k') \rightarrow (a,b,i,j,k) \rightarrow (a,d,i,j,k)_Y \rightarrow (d,c,i,j,k) \rightarrow (d,c,b')$ is a path of length~4 in~$G$.

Vertices $(a,b,i,j,k) \in \ab$, $(a,d,i,j,k)_Y \in \ady$, $(d,c,i,j,k) \in \dc$ exist since $a[i] = a[j] = a[k] = 1 = d[i] = d[j] = d[k]$ and $b[i] = b[j] = 1 = c[i] = c[k]$.
The arc $(a,b,i',j',k') \rightarrow (a,b,i,j,k)$ is index-switching in $\ab$.
The arcs $(a,b,i,j,k) \rightarrow (a,d,i,j,k)_Y \rightarrow (d,c,i,j,k)$ are two regular arcs present in $G$.
Finally the arc $(d,c,i,j,k) \rightarrow (d,c,b')$ exists since $b'[k] = c[k] = 1$.

\medskip

\textbf{AD}$\mathbf{_Y}~\mathbf{\rightarrow}~\textbf{\dcb}.$
Let $(a,d,i',j',k')_Y$ and $(d',c,b)$ be two vertices in $\ady$ and $\dcb$ respectively.
We define the indices $i := \ind(a,b,c,d')$, $j := \ind(a,c,d,d')$, and $k := \ind(a,d,d')$.
Then, $(a,d,i',j',k')_Y \rightarrow (a,d,i,j,k)_X \rightarrow (a,d',i,j,k)_Y \rightarrow (d',c,i,j,k) \rightarrow (d',c,b)$ is a path of length~4 in~$G$.

The vertices $(a,d,i,j,k)_X \in \adx$, $(a,d',i,j,k)_Y \in \ady$, $(d',c,i,j,k) \in \dc$ are present in $G$ since $a[i] = a[j] = a[k] = 1 = d'[i] = d'[j] = d'[k]$ and $d[j] = d[k] = 1 = c[i] = c[j]$.
Recall that $a[i] = a[j] = a[k] = 1 = d[j] = d[k]$ suffices for the existence of $(a,d,i,j,k)_X$ (but not for the one of $(a,d,i,j,k)_Y$).
The arc $(a,d,i',j',k')_Y \rightarrow (a,d,i,j,k)_X$ is a skew arc: it is between $\ady$ and $\adx$, and the two vector fields remain unchanged.
The arcs $(a,d,i,j,k)_X \rightarrow (a,d',i,j,k)_Y \rightarrow (d',c,i,j,k)$ are two regular arcs of $G$.
Finally the arc $(d',c,i,j,k) \rightarrow (d',c,b)$ exists since $b[i] = c[i] = 1$.

\medskip

We have proved that there is a path of length at most 4 between every (ordered) pair of vertices in~$G$, when there is no orthogonal quadruple.
Thus the diameter of $G$ is then (at most)~4.

\subsection{An orthogonal quadruple implies two vertices at distance at least 7}

We now suppose that $S$ admits at least one orthogonal quadruple, say, $a, b, c, d$.
We show that $G$ has diameter at least~7, by arguing that there is no path of length at most 6 from $(a,b,c) \in \abc$ to $(d,c,b) \in \dcb$.

The first observation is that there is no path of length at most 6 from $(a,b,c)$ to $(d,c,b)$ intersecting $\{u,v\}$.
Indeed one can check that $\dist((a,b,c),u)=4$ and $\dist(u,(d,c,b))=3$, and that $\dist((a,b,c),v)=3$ and $\dist(v,(d,c,b))=4$.
We can now rule out the existence of a path $P$ of length 6 from $(a,b,c)$ and $(d,c,b)$ in $G - \{u,v\}$.

We distinguish two cases:
\begin{itemize}
\item(a) $P$ does not intersect $\adx$, or
\item(b) $P$ intersects $\adx$.
\end{itemize}

\textbf{Case (a).}
We further distinguish two cases: either (a1) $P$ contains no index-switching arc, or (a2) $P$ contains at least one index-switching arc.
In case (a1), the three index fields cannot change at all in $P$ (recall that the skew edges are between $\adx$ and $\ady$).
Thus the first and penultimate vertices of $P$ are $(a,b,i,j,k) \in \ab$ and $(d,c,i,j,k) \in \dc$ for some $i, j, k \in [\ell]$.
The existence of these vertices imply that $a[i] = a[j] = a[k] = 1 = d[i] = d[j] = d[k]$, and vectors $b$ and $c$ both take value 1 on at least two indices among $\{i,j,k\}$.
Therefore there exists an index $h \in \{i,j,k\}$ such that $a[h] = b[h] = c[h] = d[h] = 1$.
This contradicts the fact that $a, b, c, d$ are orthogonal.

We now tackle case (a2).
Since the removal of $\ady$ separates $\abc \cup \ab$ from $\dcb \cup \dc$ in $G - \{u,v\}$, $G[\ady]$ is an independent set, and there are no edges between $\ady$ and $\abc \cup \dcb$, such a path $P$ has to contain a subpath $x \rightarrow y \rightarrow z$ with $x \in \ab$, $y \in \ady$, and $z \in \dc$.
As $P$ contains at least one index-switching arc, it cannot also contains a back-and-forth along $\ab \rightarrow \abc \rightarrow \ab$, $\ady \rightarrow \ab \rightarrow \ady$, $\dc \rightarrow \ady \rightarrow \dc$, or $\dcb \rightarrow \dc \rightarrow \dcb$.
Indeed that would amount to at least three additional arcs (at least one index-switching plus two for the back-and-forth) to the mandatory four arcs $\abc \rightarrow \ab \rightarrow \ady \rightarrow \dc \rightarrow \dcb$, hence a path of length at least 7.
Therefore, there are three indices $i, j, k \in [\ell]$ such that $x=(a,b,i,j,k)$, $y=(a,d,i,j,k)_Y$, and $z=(d,c,i,j,k)$.
Indeed every path in $G[\abc \cup \ab]$ and every path in $G[\dcb \cup \dc]$ preserve the first two vector fields.
Again the existence of $(a,b,i,j,k)$ (in $\ab$) and $(d,c,i,j,k)$ (in $\dc$) contradicts that $a, b, c, d$ are orthogonal.

\medskip

\textbf{Case (b).}
We can now assume that $P$ intersects $\adx$.
Thus $P$ has length exactly 6 and is of the form $(a,b,c) \in \abc \rightarrow \ab \rightarrow \ady \rightarrow \adx \rightarrow \ady \rightarrow \dc \rightarrow \dcb \ni (d,c,b)$.
In particular, $P$ cannot contain an index-switching edge.
If $P$ contains no skew edge too, the index fields cannot change.
So the second and sixth vertices of $P$ are some $(a,b,i,j,k) \in \ab$ and $(d,c,i,j,k) \in \dc$, and we can conclude as in case (a).

Thus $P$ has to contain at least one skew edge.
Let us show that $P$ has to contain exactly one skew edge.
We recall that the skew edges are only present between $\adx$ and $\ady$.

We first argue that the third vertex of $P$ is $(a,d',i,j,k)_Y$ for some $d' \neq d \in D$ and $i, j, k \in [\ell]$.
The first vector field cannot change in a path of the form $\abc \rightarrow \ab \rightarrow \ady$, so we only have to show that $d'$ cannot be equal to $d$.
Indeed, otherwise $a[i] = a[j] = a[k] = 1 = d[i] = d[j] = d[k]$ by the existence of $(a,d,i,j,k)_Y$.
Furthermore, the existence of the arc $(a,b,c) \rightarrow (a,b,i,j,k)$ (which has to be the first arc of $P$) implies that there is an index $h \in \{i,j,k\}$ such that $b[h] = c[h] = 1$.
This index thus contradicts the orthogonality of $a, b, c, d$.

As the third vertex of $P$ is $(a,d',i,j,k)_Y$ with $d' \neq d$, two skew edges $\ady \rightarrow \adx \rightarrow \ady$ would lead to a vertex $(a,d',i',j',k')_Y$.
This latter vertex is linked in $\dc$ to vertices of the form $(d',c',i',j',k')$ (where $c'$ can be $c$).
This cannot lead to $(d,c,b)$ since the first vector field does not change in an arc from $\dc$ to $\dcb$.

We have established that from vertex $(a,d',i,j,k)_Y$, $P$ takes exactly one skew edge, either from $\ady$ to $\adx$, or from $\adx$ to $\ady$.
In both cases, by the previous remark, the second vector field of the fifth vertex of $P$ should be $d$.
This implies that the fifth vertex of $P$ is of the form $(a,d,i',j',k')_Y$ for some indices $i', j', k' \in [\ell]$.
Indeed the skew edge of $P$ in $\ady \rightarrow \adx \rightarrow \ady$ preserves both vector fields, whereas the regular edge of $P$ in $\ady \rightarrow \adx \rightarrow \ady$ can only change one vector field, and has to change $d'$ ($\neq d$) to $d$.

The end of $P$ is thus $(a,d,i',j',k')_Y \rightarrow (d,c,i',j',k') \rightarrow (d,c,b)$, since the first two vector fields cannot be changed by an arc from $\dc$ to $\dcb$.
The existence of vertex $(a,d,i',j',k')_Y$ implies that $a[i'] = a[j'] = a[k'] = 1 = d[i'] = d[j'] = d[k']$.
The existence of the arc $(d,c,i',j',k') \rightarrow (d,c,b)$ implies that there is an index $h \in \{i',j',k'\}$ such that $c[h] = b[h] = 1$.
This yields $a[h] = b[h] = c[h] = d[h] = 1$, contradicting the orthogonality of $a, b, c, d$.

\medskip

We have ruled out the existence of a path in $G$ of length at most 6 between $(a,b,c)$ and $(d,c,b)$, when $a, b, c, d$ are orthogonal.
Hence the diameter of $G$ is at least 7 when there is an orthogonal quadruple.

\end{document}